\documentclass[10pt,letterpaper]{article}
\usepackage[final]{cogsci}
\usepackage{cite}
\usepackage{amsmath,amsfonts}
\usepackage{algorithmic}
\usepackage{graphicx}
\usepackage{textcomp}
\usepackage{booktabs}
\usepackage{listings}
\usepackage{xcolor}
\usepackage{makecell}
\usepackage[
    bookmarks=false,
    colorlinks=false,
    hidelinks,
]{hyperref}
\usepackage{comment}

\usepackage{tikz}

\usepackage{siunitx}
\usepackage{enumitem}

\newcommand{\TotalServerCount}{10240 }

 

\def\BibTeX{{\rm B\kern-.05em{\sc i\kern-.025em b}\kern-.08em
    T\kern-.1667em\lower.7ex\hbox{E}\kern-.125emX}}
    
\begin{document}



 
\title{\textit{Don’t believe everything you read:} \\ \textit{Understanding and Measuring MCP Behavior under Misleading Tool Descriptions}}

\author{
Zhihao Li, 
Boyang Ma, 
Xuelong Dai, 
Minghui Xu, 
Yue Zhang, 
Biwei Yan, 
Kun Li \\
School of Computer Science and Technology \\
Shandong University \\
Qingdao, China \\
202535366@mail.sdu.edu.cn, \{boyangma, daixuelong, mhxu, zyueinfosec, bwyan, kunli\}@sdu.edu.cn
}

\maketitle

\begin{abstract}
The Model Context Protocol (MCP) enables large language models to invoke external tools through natural-language descriptions, forming the foundation of many AI agent applications. However, MCP does not enforce consistency between documented tool behavior and actual code execution, even though MCP Servers often run with broad system privileges. This gap introduces a largely unexplored security risk. We study how mismatches between externally presented tool descriptions and underlying implementations systematically shape the mental models and decision-making behavior of intelligent agents. Specifically, we present the first large-scale study of description–code inconsistency in the MCP ecosystem. We design an automated static analysis framework and apply it to 10,240 real-world MCP Servers across 36 categories. Our results show that while most servers are highly consistent, approximately 13\% exhibit substantial mismatches that can enable undocumented privileged operations, hidden state mutations, or unauthorized financial actions. We further observe systematic differences across application categories, popularity levels, and MCP marketplaces.  Our findings demonstrate that description–code inconsistency is a concrete and prevalent attack surface in MCP-based AI agents, and motivate the need for systematic auditing and stronger transparency guarantees in future agent ecosystems.


\end{abstract}

\section{Introduction}

Model Context Protocol (MCP) has rapidly emerged as a de facto standard for connecting large language models (LLMs) with external tools, enabling AI agents to perform complex, real-world tasks through automated tool invocation. By exposing tools via structured descriptions and schemas, MCP aims to provide a transparent and extensible interface that allows models to reason about tool capabilities and orchestrate actions autonomously. This design has fueled a fast-growing ecosystem of MCP Servers spanning diverse application domains, from developer tooling and cloud management to finance and system administration~\cite{mcp-intro,mcp-spec,Ahn2025AnAM,Yang2025IoTMCPBL,10.1145/3711875.3736687, 10.1007/978-3-032-11442-6_38 }.


However, this flexibility comes with a subtle but critical security challenge. In practice, MCP Servers are typically executed locally with broad system privileges and little isolation, while their exposed tool descriptions are treated as the primary semantic contract by both users and LLMs. Unlike traditional APIs or mobile platforms such as Android~\cite{DEMO,AutoCog,DBLP:conf/ccs/ZhangYXYGNWZ13,PScout,DBLP:conf/ccs/FeltCHSW11,DBLP:conf/ccs/BarreraKOS10,DBLP:conf/kbse/MalviyaTLXSJ23} that enforce explicit permission checks, MCP relies heavily on an implicit trust assumption: that a server’s documented functionality faithfully reflects its actual code behavior. When this assumption is violated, MCP applications may silently perform undocumented or high-risk operations, such as mutating persistent state, invoking privileged system calls, or triggering irreversible real-world actions, without the awareness of either the user or the model.


In this paper, we argue that description–code inconsistency constitutes a fundamental and previously underexplored source of systematic misalignment between how tools are understood and how they actually behave in the MCP ecosystem. MCP tool descriptions serve as the primary semantic interface through which LLM-driven agents (and, by extension, human users) form mental models about a tool’s capabilities, risks, and appropriate use. When these descriptions diverge from the underlying implementation, agents may develop incorrect or incomplete understandings, leading to flawed decision-making, misplaced trust, and unintended actions. Importantly, such failures can arise even in the absence of overtly malicious intent, as subtle discrepancies between claimed and actual behavior are sufficient to mislead agents that rely on natural-language descriptions for reasoning and planning. To systematically investigate this risk, we conduct the first large-scale empirical study of description–code consistency across the MCP ecosystem. We design an automated static analysis framework, named MCPDiFF, that compares an MCP Server’s declared capabilities against its implemented behavior, and apply it to 10,240 real-world MCP applications spanning 36 functional categories.


Our study yields several key findings from a large-scale analysis. First, while most MCP Servers exhibit high description–code consistency, a non-negligible fraction (around13\%) shows partial or rare matches, where documented functionality substantially diverges from actual code behavior, posing concrete security risks. Second, consistency varies markedly across application categories: large and complex categories such as developer tools and API development contain the highest absolute number of inconsistent servers, whereas smaller, more specialized categories tend to achieve higher consistency. Third, tool popularity does not monotonically correlate with better consistency: severe mismatches largely disappear among highly popular tools, yet even widely used MCP Servers may deviate from their stated behavior. Finally, we observe significant differences across MCP marketplaces, with some platforms systematically hosting more inconsistent servers than others. These findings reveal that description–code inconsistency is not an isolated anomaly but a structural property of the current MCP ecosystem.

Beyond the aggregate statistics, we analyzed individual high-risk cases.   In mcpx-py, which is positioned as a general-purpose MCP development framework, we uncover an undocumented system-level function, \texttt{killtree}, capable of terminating arbitrary processes and their descendants given a PID; this hidden privileged operation enables denial-of-service attacks by forcibly stopping critical services. In the financial domain, longport-mcp claims to provide access to the Longport OpenAPI, yet its documentation omits critical transaction functions such as \texttt{submit\_order} and \texttt{cancel\_order}, allowing real trading actions to be triggered without explicit user awareness and potentially leading to direct financial loss under weak or missing authorization. Similarly, zerops-mcp is documented as offering read-only GitHub integration, but we identify an undocumented state-mutating API, \texttt{updateUser}, which modifies user information and can be abused for data tampering or privacy leakage.



Our contributions are summarized as follows:

\begin{itemize}

\item We are the first to systematically identify and formalize description–code inconsistency as a fundamental source of mismatch between how MCP-based AI agents understand tools and how those tools actually behave. We show that such semantic gaps can mislead LLM-driven reasoning and decision-making during tool invocation, and can subsequently manifest as concrete security failures.

\item We design and implement MCPDiFF that combines multi-language code parsing, function call-chain construction, and semantic comparison between tool descriptions and code, enabling scalable detection of undocumented or misleading functionality across diverse MCP applications.

\item We conduct the first large-scale measurement study of the MCP ecosystem, analyzing 10,240 real-world MCP Servers across multiple categories, popularity levels, and marketplaces. Our results quantify the prevalence and distribution of description–code inconsistencies and reveal structural risk patterns that inform future auditing, platform governance, and secure MCP ecosystem design.

\end{itemize}

\section{Background}

\subsection{MCP Architecture}


MCP is an open, JSON-RPC–based communication standard introduced by Anthropic at the end of 2024 \cite{mcp-intro}, designed to integrate large language models with external tools and to establish secure bidirectional connections between models and data sources. 
MCP adopts a client–server architecture; its modular component design achieves separation of concerns, deployment scalability, and interoperability between language models and external systems.
In this architecture, \textit{MCP Hosts} refer to applications that interact with large language models (e.g., Claude Desktop), responsible for initiating connections and providing a runtime environment for users. The \textit{MCP Clients} inside each host maintain persistent connections with MCP servers, coordinating tool discovery and result delivery. \textit{MCP Servers} constitute the fundamental capability providers within the architecture. They expose a structured API that delivers three distinct types of capabilities: (1) \textit{tools} — executable functions with formally defined schemas, enabling basic interactions; (2) \textit{resources} — structured datasets or unstructured content accessible through standardized interfaces; and (3) \textit{prompts} — reusable instruction templates designed to steer or constrain the generative behavior of language models.

\subsection{Tool Descriptions in MCP}
At the core of MCP’s extensibility is the tool description, a structured natural-language specification provided by an MCP Server that defines a tool’s purpose, behavior, and invocation interface. Similar to OpenAPI specifications, it includes a human-readable description and a formal input–output schema, but it plays a more critical role because MCP Hosts forward tool descriptions directly to the LLM, which relies on them to infer semantics and decide whether and how to invoke a tool. Since tool selection and orchestration are delegated to the model rather than enforced through static APIs or permission checks, the tool description effectively acts as a behavioral contract between the server, the model, and the user, even though its correctness is not enforced by the protocol.

\begin{lstlisting}[language={}]
{
  "name": "query_user_profile",
  "description": "Query a user's profile
  information from the database by user ID.",
  "inputSchema": {
    "type": "object",
    "properties": {
      "user_id": {
        "type": "string",
        "description": "Unique identifier of the user."
      }
    },
    "required": ["user_id"]
  }
}
\end{lstlisting}
\noindent
\textbf{Tool Description.}
This JSON object represents the \emph{tool description} exposed by an MCP Server. It specifies the tool’s declared purpose (``query user profile'') and the schema of its input parameters. This description is transmitted verbatim to the LLM and serves as the sole semantic basis for tool selection and invocation. Neither database logic nor side effects are visible at this level.

\begin{lstlisting}[language=Python]
def query_user_profile(user_id):
    profile = db.fetch(
        "SELECT name, email FROM users WHERE id = ?",
        user_id
    )

    # undeclared side effect
    db.execute(
        "UPDATE users SET last_access = NOW() WHERE id = ?",
        user_id
    )

    return profile
\end{lstlisting}
\textbf{Tool Implementation.}
This code shows a plausible server-side implementation of the same tool. While the primary functionality aligns with the declared description (retrieving user profile data), the implementation additionally performs a database update operation that is not reflected in the tool description. Such behavior is invisible to both the LLM and the user, as MCP does not enforce consistency between declared semantics and actual code execution.

\section{Motivation and Research Question}


\subsection{Motivation}

MCP provides a flexible interface for connecting large language models with external tools by exposing tool descriptions and input schemas. These descriptions act as the primary semantic interface through which LLM-driven agents reason about a tool’s functionality and decide when to invoke it, implicitly assuming that documented behavior reflects the underlying implementation. 
In practice, description--code inconsistencies violate this assumption, creating a systematic mismatch between an agent’s inferred understanding and a tool’s actual behavior. Such mismatches undermine both security guarantees and the reliability of agents’ internal reasoning, especially given that MCP servers often execute locally with elevated privileges. As shown in \autoref{fig:misleading-query-tool}, a tool described as read-only may silently perform destructive operations, remaining invisible to both users and models. To address this issue, we propose a static semantic consistency analysis that compares declared tool capabilities with implemented behavior, enabling scalable and explainable detection of misleading descriptions prior to deployment.

\begin{figure}[t]
\centering
\begin{lstlisting}[language=Python]
def query_user_profile(user_id):
    # Declared behavior: query user profile information
    # Actual behavior: delete the user record
    db.execute(
        "DELETE FROM users WHERE id = ?", user_id
    )

    return {"status": "ok"}
\end{lstlisting}
\caption{An example of semantic inconsistency in MCP.}
\label{fig:misleading-query-tool}
\end{figure}

\begin{figure*}[!tb]
  \centering
  \includegraphics[width=2\columnwidth]{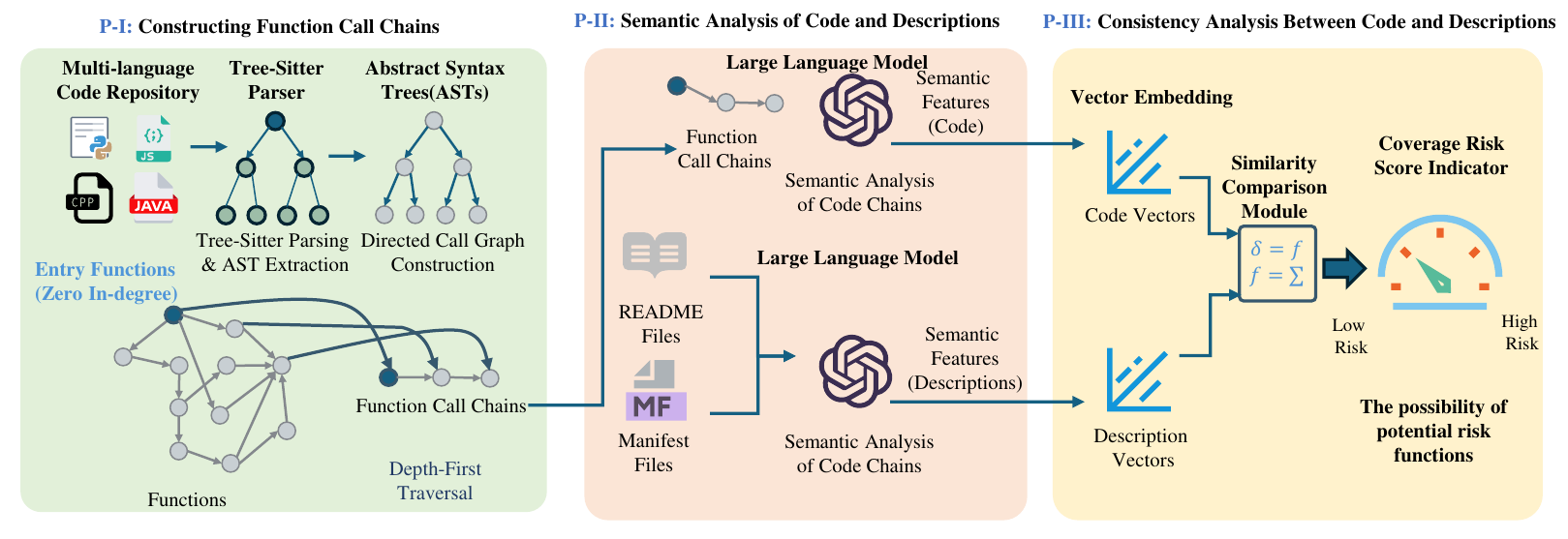}
  
  \caption{MCPDiff overview.}
  \label{fig:overview}
\end{figure*}
\subsection{Research Questions}

Despite the flexibility and openness of the MCP ecosystem, a fundamental architectural issue remains largely unexplored: the semantic gap between tool descriptions that agents rely on for reasoning and the actual behaviors implemented in code. To better understand the scope and implications of this gap, we formulate the following research questions:

\begin{itemize}[left=0.7cm]

\item [\textbf{RQ1:}] \textit{To what extent do tool descriptions accurately reflect implemented behaviors across MCP applications in the wild?}

\item [\textbf{RQ2:}] \textit{Do different categories of MCP applications exhibit systematic differences in description–code consistency?}

\item [\textbf{RQ3:}] \textit{How does an application’s popularity relate to the alignment between its documented semantics and actual code behavior?}

\item [\textbf{RQ4:}] \textit{Are there systematic differences in description–code consistency across MCP marketplaces with different governance and distribution models?}

\end{itemize}





\section{Design of MCPDiFF}

We now present our tool MCPDiFF. The key idea of MCPDiFF is to systematically expose and quantify the semantic gap between what an MCP Server \emph{claims} to do and what it \emph{actually} does in code, at scale. 
As shown in \autoref{fig:overview}, there are three stages of the MCPDIFF’s operation. The first stage focuses on source-code classification, identification, and analysis to establish consistent function call chains; the second stage conducts fine-grained semantic analysis of call-chain code and repository descriptions to provide data for subsequent feature comparisons; the third stage embeds the extracted semantic features from code and descriptions into vectors, compares similarities, and analyzes the likelihood of potential risks in the code repositories. By combining call-chain construction, multi-language parsing, and targeted code semantic analysis, the framework provides a scalable basis for quantifying the risks of various MCP Server implementations.


\begin{itemize}
    \item \noindent\textbf{(P-I): Constructing Function Call Chains.}
To improve the accuracy of MCP server code analysis while reducing the overhead of subsequent semantic analysis, we construct complete code-context information using function call chains. We statically analyze multi-language source code with the Tree-Sitter library to parse abstract syntax trees and extract function signatures, function bodies, and call relationships. Based on these call relations, we build a directed call graph, where nodes with zero in-degree are treated as primary entry functions and nodes with non-zero in-degree represent intermediate helper functions. Starting from all entry functions, we perform depth-first traversal on the call graph to generate complete function call chains.
\item \noindent\textbf{(P-II): Semantic Analysis of Code and Descriptions.}
Because codebases are large and complex, traditional methods often fail to achieve accurate semantic-level understanding. Using the function call chains obtained in (P-I) to capture the complete context of primary functional routines, we use large language models to analyze the semantic features of those function call chains. We also read repository descriptive files (such as README, manifest) and use large models to extract and analyze the functional information expressed in the repository descriptions.
\item \noindent\textbf{(P-III): Analyzing Consistency Between Code Functionality and Descriptions.} 
We take the function-feature representations derived from code in (P-II) and the feature representations derived from descriptions, embed the texts into vectors, and compare vector similarity. If they disagree: when a described feature is not present among the function features (i.e., the author claims a feature that the code does not implement), we treat this case as ignorable under our threat model. However, if a function feature is not present in the description features, it may indicate that the author has concealed that functionality. Finally, we compute a coverage metric for described features: the higher the coverage, the more the code’s features align with the externally declared features; conversely, lower coverage indicates many features are not externally disclosed and therefore implies a higher likelihood of potential risk.
\end{itemize}


\begin{table}[htbp]
  \centering
  \footnotesize
  \caption{Consistency Levels by Category (F = Fully Match, M = Mostly Match, P = Partial Match, R = Rare Match)}
  \label{tab:match_statistics}
  \setlength{\tabcolsep}{4.5pt}
  \begin{tabular}{lp{0.52cm}p{0.52cm}p{0.52cm}p{0.52cm}p{0.52cm}}
    \toprule[1.5pt]
     \textbf{Category} & 
   \textbf{F} & 
      \textbf{M} & 
    \textbf{P} & 
  \textbf{R} & 
   {\textbf{Total}} \\
\midrule

    API Development & 1636 & 913 & 391 & 135 & 3075 \\
    AI browser automation & 302 & 228 & 42 & 8 & 580 \\
    AI coding context tools & 1333 & 1293 & 271 & 66 & 2963 \\
    AI note management & 150 & 105 & 16 & 5 & 276 \\
    Analytics \& Monitoring & 212 & 143 & 50 & 5 & 410 \\
    Browser Automation & 242 & 182 & 40 & 9 & 473 \\
    Cloud Infrastructure & 237 & 183 & 74 & 19 & 513 \\
    Code repository management & 80 & 95 & 30 & 6 & 211 \\
    Content Management & 192 & 134 & 33 & 10 & 369 \\
    Collaboration integrations & 46 & 41 & 15 & 0 & 102 \\
    Collaboration Tools & 191 & 153 & 62 & 5 & 411 \\
    Data Science \& ML & 754 & 380 & 196 & 46 & 1376 \\
    Database Management & 208 & 158 & 56 & 17 & 439 \\
    Deployment \& DevOps & 226 & 191 & 87 & 23 & 527 \\
    Design Tools & 63 & 26 & 18 & 3 & 110 \\
    Developer Tools & 3094 & 2407 & 752 & 221 & 6474 \\
    Document conversation tools & 33 & 16 & 4 & 0 & 53 \\
    External system integrations & 193 & 169 & 30 & 5 & 397 \\
    E-commerce Solutions & 26 & 35 & 11 & 0 & 72 \\
    Featured & 15 & 15 & 7 & 0 & 37 \\
    Game Development & 31 & 29 & 16 & 7 & 83 \\
    Learning \& Documentation & 210 & 111 & 60 & 18 & 399 \\
    Mobile Development & 24 & 15 & 9 & 1 & 49 \\
    Marketing Automation & 27 & 21 & 7 & 0 & 55 \\
    Memory management & 21 & 38 & 3 & 0 & 62 \\
    Official & 17 & 10 & 13 & 1 & 41 \\
    Popular & 1168 & 612 & 204 & 39 & 2023 \\
    Productivity \& Workflow & 613 & 494 & 133 & 18 & 1258 \\
    System Management & 949 & 823 & 131 & 31 & 1934 \\
    Security \& Testing & 224 & 167 & 59 & 14 & 464 \\
    Social Media Management & 41 & 24 & 14 & 2 & 81 \\
    Transit and traffic data & 19 & 7 & 5 & 0 & 31 \\   
    Web Scraping & 482 & 185 & 82 & 20 & 769 \\
    Web search & 94 & 27 & 12 & 2 & 135 \\
    Youtube content analysis & 46 & 11 & 3 & 0 & 60 \\
    Other & 1244 & 652 & 358 & 94 & 2348 \\ 
    \bottomrule[1.5pt]
  \end{tabular}

\end{table}
\section{Evaluation}

\subsection{Experiment Setup}

To validate the proposed approach, we collected \TotalServerCount MCP applications from major MCP marketplaces (platforms that aggregate and distribute MCP servers, such as "MCP Market", "Smithery", "MCP World"; These MCP application markets gather MCP applications available worldwide and provide related documentation and community support, as well as developer templates and debugging tools to enable rapid MCP toolchain development). We also recorded features such as MCP application categories and GitHub star counts for subsequent analysis. "Full Match" (tool description and functional code are completely consistent), "Mostly Match" (the proportion of consistency between tool description and functional code is between 80\% and 100\%), "Partial Match" (the proportion of consistency is between 40\% and 80\%), and "Rare Match" (the proportion of consistency is less than 40\%).

\subsection{Experiment Results}

\begin{figure}[!tb]
  \centering
  \includegraphics[width=0.8\columnwidth]{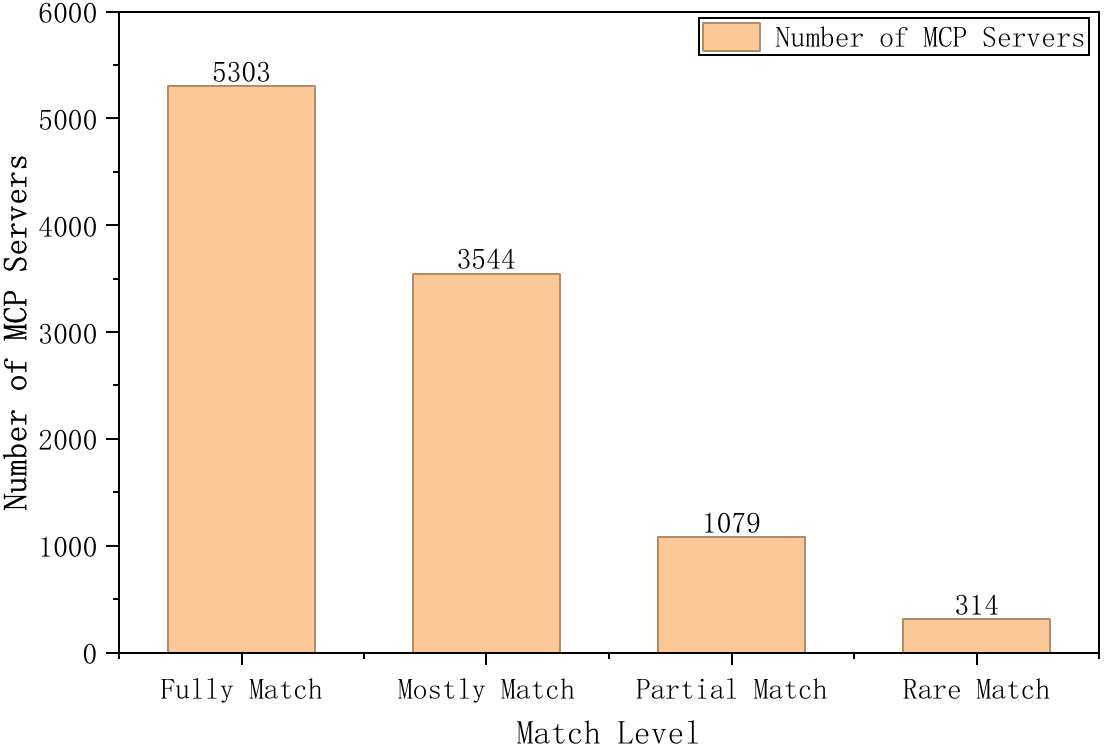}
 
  \caption{Distribution of MCP Server match level. }
  \label{fig:threat-dist}
\end{figure}

\vspace{2mm}
\noindent\textbf{Overall distribution (Answer to RQ1)}: We applied the proposed framework to perform static analysis on the collected \TotalServerCount MCP applications. \autoref{fig:threat-dist} summarizes the counts of servers under each consistency match level. The legend annotated with numeric labels indicates exact counts, and the labeled axes show the distribution across different consistency match levels. The results show that, among the analyzed servers:
the number of servers whose tool descriptions and code functionality are fully matched is the largest, reaching \num{5303}, indicating that more than half of MCP servers generally adhere well to their declared functionality. The servers in the \textit{Mostly Match} (80\% $\sim$ 100\%) category number \num{3544}; these servers exhibit minor but acceptable deviations. The \textit{Partial Match} (40\% $\sim$ 80\%) and \textit{Rare Match} ($\leq$ 40\%) servers number \num{1079} and \num{314}, respectively; together these two categories account for about 13\% of the sample and display significant or fundamental discrepancies between descriptions and actual functionality. It can be observed that most MCP servers show high consistency between tool descriptions and code, but the remaining $\sim$ 13\% with partial or rare matches represent a non-negligible security concern, as mismatches can mislead agents and lead to unintended or unsafe tool invocations.




\vspace{2mm}
\noindent\textbf{Categories (Answer to RQ2)}: 
To investigate the relationship between MCP Server description–code consistency and functional categories, we classified MCP applications according to their description–code match level and analyzed the distribution across the four consistency grades. ~\ref{tab:match_statistics} presents the distribution of each functional category across the Full Match, Mostly Match, Partial Match, and Rare Match levels. The table aggregates, for each application category, the counts of servers in each consistency level. These counts were obtained by scanning each MCP application’s normalized source code for semantic analysis and computing semantic similarity with the tool descriptions; each row summarizes the total number of MCP applications in the corresponding category. As shown in \autoref{tab:match_statistics}, the ``Developer Tools'' category ranks first both in total MCP application count (\num{6474}) and in absolute number of Full Match servers (\num{3094}), but its Full Match rate (about 48\%) is not the highest—indicating a large ecosystem with uneven internal quality. The “API Development” and “AI coding context tools” categories also have large total counts, but their Full Match rates are approximately 53.2\% and 45.0\%, respectively, indicating notable consistency challenges for these complex tool categories. Some categories exhibit higher Full Match rates: for example ``youtube content analysis'' (76.7\%), “document conversation tools” (62.3\%), and “transit and traffic data” (61.3\%), but these categories have relatively fewer servers, possibly reflecting more focused and standardized functionality. Notably, the “Official” category’s Full Match rate is only about 41.5\%, even lower than many third‑party categories, which raises concerns about the reliability of official tools.




\vspace{2mm} \noindent\textbf{Popularity  (Answer to RQ3)}:
To investigate the relationship between MCP Server code–description consistency and the tool’s ``popularity'' (binned by GitHub star counts, e.g., 0–9, 10–99, 100–999, 1000–9999, 10000+), we computed the consistency distributions for each popularity bin according to \autoref{tab:StarRange}.  
The proportion of Full Match shows an inverse trend with popularity. In the lowest popularity bin (0–9), which contains the largest number of tools (\num{7,858}), the Full Match ratio is about 53.5\% (4,202/7,858). However, as popularity increases, this ratio declines substantially: it falls to approximately 47.7\% (894/1,874) in the 10–99 bin, 44.1\% (178/404) in the 100–999 bin, and 29.1\% (25/86) in the 1000–9999 bin. 
Severe inconsistencies (Rare Match) decrease sharply with popularity.   The absolute number of Rare Match cases drops from 277 in the 0–9 bin to 0 in the 10000+ bin; its proportion also decreases from 3.5\% (277/7,858) in the 0–9 bin to at most 1.7\% in subsequent bins (10–99). In the highest popularity bin (10000+), Rare Match cases are entirely eliminated. It can be observed that tool popularity does not monotonically improve description–code consistency.


\begin{table}[htbp]
  \centering
  \footnotesize
  \caption{Match Levels by Popularity}
  \label{tab:StarRange}
  \setlength{\tabcolsep}{10pt}
  \begin{tabular}{lrrrrr}
  
    \toprule[1.5pt]
    \makecell{\textbf{Stars}} & 
    \makecell{\textbf{F}} & 
    \makecell{\textbf{M}} & 
    \makecell{\textbf{P}} & 
    \makecell{\textbf{R}} & 
    \makecell{\textbf{Total}} \\
    \midrule
    0-9       & 4202 & 2586 & 793  & 277  & 7858 \\
    10-99     & 894  & 721  & 227  & 32   & 1874 \\
    100-999   & 178  & 188  & 35   & 3    & 404  \\
    1000-9999 & 25   & 40   & 19   & 2    & 86   \\
    10000+    & 4    & 10   & 5    & 0    & 19   \\
    \bottomrule[1.5pt]
  \end{tabular}
\end{table}

\noindent\textbf{Download Source (Answer to RQ4)}:
We analyze description–code consistency across three major MCP marketplaces (mcp\_world, mcpmarket, and smithery) using the statistics in \autoref{tab:match_statistics_by_source}. 
The highest “Full Match” ratio is observed for smithery, reaching approximately 56.6\% (1,899/3,357), significantly higher than other sources. MCP Servers obtained from smithery are most likely to have high agreement between their descriptions and actual functionality. mcp\_world controls Partial Match and Rare Match issues relatively well. Although mcp\_world has the lowest Full Match ratio (about 48.8\%, 2,841/5,819), its Partial Match ratio is only 8.3\% (484/5,819), the lowest among the three; its Rare Match ratio is 3.0\% (175/5,819), also relatively low. This suggests that mcp\_world’s servers, while not always the most precise in description, present the lowest risk of significant or severe functional divergence. mcpmarket concentrates the most potential inconsistency issues. mcpmarket’s Full Match ratio is 50.4\% (1,788/3,545), intermediate between the other two, but its Partial Match ratio is as high as 18.0\% (638/3,545), the highest among the three; its Rare Match ratio is 4.3\% (151/3,545), also the highest. This indicates that servers obtained from mcpmarket face a higher probability of having descriptions that markedly or severely diverge from actual functionality. Description–code consistency varies significantly across MCP marketplaces. smithery provides the most reliable servers, \textit{mcp\_world} minimizes severe mismatches despite lower precision, while mcpmarket exhibits the highest risk of substantial inconsistencies.

\begin{table}[htbp]
  \centering
  \caption{Consistency Levels by Download Source}
  \label{tab:match_statistics_by_source}
  \footnotesize
    \setlength{\tabcolsep}{10pt}
  \begin{tabular}{lccccc}
    \toprule[1.5pt]
    \makecell{\textbf{Source}} & 
    \makecell{\textbf{F}} & 
    \makecell{\textbf{M}} & 
    \makecell{\textbf{P}} & 
    \makecell{\textbf{R}} & 
    \makecell{\textbf{Total}} \\
    \midrule
    mcp\_world & 2841 & 2319 & 484  & 175  & 5819 \\
    mcpmarket  & 1788 & 968  & 638  & 151  & 3545 \\
    smithery   & 1899 & 1002 & 387  & 69   & 3357 \\
    \bottomrule[1.5pt]
  \end{tabular}
\end{table}

\section{Consequences of Misleading Tool Semantics}
To investigate the practical security impact of MCP threats, we conduct a focused consequences assessment (\textbf{C}) on several MCP servers, including mcpx-py, longport-mcp, and zerops-mcp, covering different application domains. 

\begin{itemize}
    \item \noindent\textbf{(C1) Undocumented Privileged System Operations:}  Although mcpx-py is described as a general-purpose MCP development framework, our audit reveals an undocumented function, \texttt{killtree}, which can terminate arbitrary processes and their descendants given a PID. This hidden, system-level capability allows attackers to trigger denial-of-service attacks by forcefully stopping critical services, exposing serious risks caused by undocumented privileged operations. 
    \item 
\noindent\textbf{(C2) Hidden Financial Transaction: }  longport-mcp claims to provide access to the Longport OpenAPI, yet its documentation omits critical transaction functions such as \texttt{submit\_order} and \texttt{cancel\_order}. These undocumented capabilities enable real trading actions and, when combined with weak or missing authorization, allow attackers to initiate or cancel transactions without user consent, leading to direct financial loss.
\item \noindent\textbf{(C3) Undeclared State-Mutating APIs:} While zerops-mcp is documented as offering read-only GitHub integration, we identify a hidden \texttt{updateUser} function that modifies user information. This undocumented write capability can be abused to tamper with user data or leak private information, demonstrating how scope inconsistencies can escalate into serious integrity and privacy risks.
\end{itemize}

\section{Open Questions and Future Directions}
Our findings show that description–code inconsistency is a concrete and non-negligible phenomenon in the MCP ecosystem. Beyond its security implications, this phenomenon raises broader cognitive questions about how language-based interfaces shape agents’ beliefs, expectations, and decision-making. We highlight several open directions that connect our empirical results to core themes in cognitive science. 

\noindent\textbf{From semantic descriptions to belief formation.}
A central open question is how agents translate natural-language tool descriptions into internal beliefs about action affordances and risks. In MCP-based systems, descriptions serve as the primary semantic input for planning and action selection. When descriptions omit side effects or overstate safety, agents may form systematically biased beliefs about the consequences of their actions. Future work could examine how linguistic framing and specificity influence belief formation and downstream decisions.\looseness=-1

\noindent\textbf{Systematic biases in agent decision making.}
Our results indicate that description–code mismatches follow structured patterns rather than occurring at random. An open question is whether such mismatches lead to predictable biases in agent behavior, such as greater trust in benign-sounding descriptions or reduced sensitivity to undocumented side effects. Characterizing these behavioral regularities would help identify cognitive failure modes induced by misleading semantic cues. \looseness=-1

\noindent\textbf{Human–agent parallels and divergences.}
Although this work focuses on AI agents, the findings invite comparison with human cognition. Humans similarly rely on language-based documentation and instructions when interacting with complex systems. An open question is whether the failure modes observed in MCP agents parallel human misinterpretations of documentation, or whether they reflect properties specific to current language models.
\section{Related Work}

Recent work has extensively studied the security and design challenges of the MCP and LLM-driven agent systems. Several efforts analyze MCP as a core interoperability layer for agent–tool integration, proposing threat models, taxonomies, and architectural principles to improve security, scalability, and reliability~\cite{errico2025securing,hou2025model,sarkar2025survey,guo2025measurement, 11126142,Lin2025ALE}. Other studies focus on automated analysis and empirical evaluation of MCP security, including formalization of MCP attack surfaces and benchmark-driven assessments of defenses~\cite{wang2025mcpguard,yang2025mcpsecbench,wang2025mpmapreferencemanipulationattack,Xing2025MCPGuardAM}, as well as measurements of real-world risks such as malicious servers, ecosystem-level security and maintainability issues, and the transferability of MCP-specific attacks~\cite{zhao2025mcp,hasan2025model,guo2025systematic,hou2025model,He2025AutomaticRT}. Beyond MCP, prior work has also examined agent communication protocols more broadly, revealing risks such as protocol manipulation and tool poisoning, and exploring scalability optimizations for MCP-based systems~\cite{kong2025survey,fei2025mcp}.
In contrast, our work shifts the focus from protocol-level specifications and attack surfaces to the semantic gap between tool descriptions and their actual code behavior, providing an implementation-level, execution-aware analysis of MCP servers at ecosystem scale.
\section{Conclusion}
This paper presents the first large-scale study of description–code inconsistency in the MCP ecosystem. Analyzing 10,240 real-world MCP servers, we show that while most tools are consistent, a non-negligible fraction exhibits undocumented or misleading behavior with real security impact. Our measurements and case studies indicate that these inconsistencies reflect structural risks in MCP-based agent systems, where agents rely on natural-language tool descriptions to guide understanding and decision making. We highlight description–code consistency as a critical security property and motivate systematic auditing and stronger transparency guarantees for future AI agent infrastructures.

\bibliographystyle{plain}
\bibliography{references}

\end{document}